\documentclass{aa}
\usepackage{txfonts}
\usepackage{graphicx}
\usepackage{natbib}
\usepackage{subfigure}

\begin{document}
\title{Revised Results for Non-thermal Recombination Flare Hard X-Ray Emission}
\author{J.~C. Brown\inst{1} \and  P.~C.~V. Mallik \inst{1} \and N.~R. Badnell \inst{2}}

\institute{Department of Physics and Astronomy, University of
Glasgow, Glasgow G12 8QQ, U.K. \and Department of Physics, University of Strathclyde, Glasgow G4 0NG, U.K.}

\offprints{P.C.V. Mallik, \email{pmallik@astro.gla.ac.uk}}

\authorrunning{Brown, Mallik \& Badnell}
\titlerunning{Non-thermal recombination revisited}
\date{\today}

\abstract
{Brown and Mallik (BM) recently showed that, for hot sources, recombination of non-thermal electrons (NTR) onto highly
   ionised heavy ions is not negligible compared to non-thermal
   bremsstrahlung (NTB) as a source of flare hard X-rays (HXRs) and so
   should be included in modelling non-thermal HXR flare emission. They further claimed that, in some cases, NTR can be much larger than NTB with important consequences for flare physics.}
   {In view of major discrepancies between BM results for the {{\it thermal}}
   continua and those of the Chianti code and of RHESSI solar data, we
   critically re-examine and correct the BM analysis and modify the
   conclusions concerning the importance of NTR.}
   {The BM and Chianti element abundances and ionisation
   fractions as a function of temperature $T$ for relevant elements are
   found to agree well. Although the analytic Kramers expression used by BM is correct for the purely hydrogenic recombination cross section, the heuristic expressions used by BM to extend the Kramers expression beyond the `bare nucleus' case to which it applies had serious errors. BM results have therefore been recalculated using corrected expressions, which have been validated against the results of detailed calculations.}
   {The BM results are found to be correct for NTR onto Fe 26+ and a factor of 2 too high for Fe 25+. Thus, at high enough $T >40$MK for these to exist, such NTR strongly dominates NTB in the deka-keV range just as BM claimed. However, at such $T$, thermal continuum
   dominates NTR and NTB in this energy range unless the non-thermal
   electron density is a very large fraction, $f_c$, of the total as in some
   coronal HXR sources. At $T\approx 10-30$ MK the dominant ions are Fe 22+,
   23+, 24+ for which BM erroneously overestimated NTR emission by around an
   order of magnitude. Contrary to the BM claim, NTR in hot flare plasmas does {{\it not}} dominate over NTB, although in some cases it can be comparable and so still very important in inversions of photon spectra to derive electron spectra, especially as NTR includes sharp edge features.}
   {The BM claim of dominance of NTR over NTB in deka-keV emission is incorrect due to a serious error in their analysis. However, the NTR contribution can still be large enough to demand inclusion in spectral fitting, the spectral edges having potentially serious effects on inversion of HXR spectra
   to infer fast electron spectra.}


\maketitle

\section{Introduction}
\cite{Brown&Mallik:2008, Brown&Mallik:2009} (BM) emphasised the importance of non-thermal recombination (NTR), previously neglected as a mechanism for production of solar hard X-rays (HXRs). BM obtained analytic Kramers approximations for recombination and bremsstrahlung continua and showed the resulting spectra from Maxwellians and cut-off power-laws to elucidate the importance of NTR. BM also used typical flare parameters to support their findings and concluded that NTR is not only an {\it important} component of the HXR spectrum in solar flares, but can be the {\it dominant} source in hot plasmas. This had serious implications for spectral inference of electron distributions and for electron number and energy budgets. However, BM did not compare their results with actual flare HXR data from instruments such as RHESSI \citep{Lin:2002} nor their thermal spectra with computed atomic/radiation databases such as Chianti \citep[e.g.][]{Dere:2009}. In this paper, we discuss such comparisons, which revealed serious discrepancies overlooked by us till now and which led us to discover blunders in BM that we amend here, with apologies.

 Chianti includes only thermal emission (bremsstrahlung (TB) and recombination (TR) plus lines) and not non-thermal bremsstrahlung (NTB) nor NTR, so we can only compare it with BM results for the Maxwellian case. Fig. \ref{chianticomp} reveals that the `exact' Chianti and the approximate BM results agree well for TB spectra but that the TR emission BM predicts is much higher around the 10 keV range due to the Fe and Ni emission in BM which shows large recombination edges (around 7-8 keV) almost invisible in Chianti. \cite{Landi:2007} emphasised the importance of TR compared to TB, but his predictions and spectra were also quite different from BM. (Unlike Chianti, BM spectra did not include emission lines but even the lines in Chianti are smaller than BM's recombination edges). After BM, we also recognised that Chianti thermal spectra can fit pretty well to what RHESSI observes, a fact pointing strongly toward some flaw in the BM modelling. These discrepancies motivated this paper which uncovers and corrects several errors in BM.

Having checked all our algebra and arithmetic, compared our Kramers purely hydrogenic expressions with those of \cite{Landi:2007} and others and checked the accuracy of Kramers' cross-sections against more exact atomic results
and also of BM against more exact Maxwellian-weighted emission rates (all good within about 15-20\% at worst), and that BM element abundances and ionisation fractions were in line with Chianti, we had to ask whether there was a more basic flaw in the BM heuristic extension of Kramers expressions for H-like ions to He- and Li-like and Li-like to Be-like. We found that there was. A crucial clue lay in a closer look at Fig. \ref{chianticomp}, which shows that the energies and sizes of edges for all the ions in the BM code agree quite well with those in Chianti, {\it except} for Fe and Ni, which proved discrepant. These are the only significant contributors which are not near full ionisation.

 Sect. 2.3 of \cite{Brown&Mallik:2008} stated ``Noting that $Q_R \propto 1/m^3$ we include here only recombination to $m=1$ (in the sense of the lowest empty level of the ion - hydrogenic with $Z=Z_{\rm eff}$ - not of the atom). Higher $m$ contributions are weaker, being $\propto 1/m^3$ though extending to lower
energies with edges at $Z_{\rm eff}^2\chi/m^2$", where $Q_R$ is the recombination cross-section. Herein lies the source of the BM blunder and the source of discrepancy. The relevant `$m$' in the cross section for the lowest unfilled level is {\it not} $m=1$ in general but the principal quantum number for that lowest unfilled level of the ion involved. For Fe26+ and 25+ $m=1$, but for Fe 24+ to Fe 17+, for example, $m\neq 1$ but  $=2$! So the BM interpretation of $m$ was at fault. Henceforth, we shall denote the principal quantum number by its standard notation `$n$' to avoid further confusion. In ionic species which already have two or more electrons present, the smallest $n$ value $n_{\rm min}$ is 2, and if there are 10 or more electrons, it is $n_{\rm min}=3$ and so on. (Recombination rates to levels with $n>n_{\rm min}$ still fall off as $1/n^3$, so are rather small in comparison). The consequences of this BM error are :

\begin{itemize}
\item since, for typical hot flare temperatures of 20-30 MK, Fe 24+ and Fe 23+ are the most abundant Fe ions,  $n_{\rm min}=2$, not $1$. Consequently, the magnitude of the recombination emission is down compared to BM by a factor around $1/n_{\rm min}^3=1/8$.

\item equally importantly, the locations of recombination edges for these Fe species are no longer at $Z_{\rm eff}^2\chi$, or $Z_{\rm eff}^2\chi + E_c$ in the presence of a low energy cut-off $E_c$, but at $Z_{\rm eff}^2\chi/4$ or $Z_{\rm eff}^2\chi/4 + E_c$ respectively.

\item a separate issue missed by BM is that the Kramers
formula assumes, and is applicable to, recombination into an
empty shell. For highly-charged ions, reasonable account is taken of
initially partially filled  $n$-shells by applying a
`vacancy factor' $p_n$ to the usual Kramers formula.
The simplest choice for $p_n$ is $N_v/N_n$ where $N_n = 2n^2$
is the total electron occupation number of an $n$-shell and $N_v$
the number unoccupied. For recombination of a H-like ion to $n=1$
of a He-like ion, $p_1=1/2$. For recombination into a partially
filled $n=2$ shell, $N_2=8$ would follow. However, a more
accurate result is obtained on recognising that recombination
into $n=2$ in the Kramers formula is dominated by the 6 $p$-states,
i.e., little of the Kramers $n=2$ result arises from
recombination into $s$-states, at least at the electron energies
of interest here. Thus, we take $N_2 = 6$ and $N_v$ the number of
unoccupied $2p$ states, i.e. $p_2 = 1$ for Li and Be-like initial ions and
$p_2 = 5/6,~ 4/6, ...~ 1/6$ for B- through F-like initial ions.
Comparisons of such modified Kramers cross sections have been made
with the results of detailed calculations using the AUTOSTRUCTURE code
\citep[cf.][]{Badnell:2006} for initial H-like through
to F-like Fe ions and agreement to within 20\% is obtained at the electron
energies of interest here.
\end{itemize}

These considerations explain why {\it all} the  elements apart from Fe and Ni produced  edge positions and strengths in BM reasonably comparable with Chianti, since those elements are almost fully ionised and $n_{\rm min}=1$ at relevant temperatures. The high $Z_{\rm eff}$ ions Fe and Ni are different since they are predominantly in ionic stages with $n=1$ filled and $n_{\rm min}=2$. (For super hot plasmas, say $>$ 40 MK, Fe 25+ and Fe 26+ with $n=1$ start to become important, but at those temperatures the thermal emission begins to dominate emission up to 50-60 keV and to dominate any NTR emission and recombination edges in particular, an issue we discuss later). In this paper we address the implications of these corrections. For full details see \cite{Mallik:2010}. The BM results
were mainly for the widely used, though unphysical, electron flux
spectrum $F(E) \propto E^{-\delta}$ at $E\geq E_c$ and zero below that.
For generality of our corrections here we also give the NTR expression
for general $F(E)$.

Sections 2 and 3 provide the key amended expressions and results to replace the erroneous ones in BM and new plots to emphasise the differences from BM, and to re-assess the importance of NTR. Although the magnitude of our NTR predictions has been much reduced, it is still significant and requires consideration,  especially in cases where the spectral index $\delta$ is large, the low-energy cut-off $E_c$ is small, and temperature $T$ is between 20 and 30 MK. Our overall conclusions are discussed in Section 4.

\section{Alterations to relevant expressions}

Revisions of the recombination cross-section expressions given in Eqs. (11) and (12) of \cite{Brown&Mallik:2008} have to be carried through to all other recombination expressions, e.g. Eqs. (16, 19, B.2, B.5, B.12) in \cite{Brown&Mallik:2008} and Eq. (2) in \cite{Brown&Mallik:2009}, to  show the sum over $n \geq n_{\rm min}$, with $n_{\rm min} \neq 1$ in general. So, the revised thermal recombination expression (from Eq. (B.5) of \cite{Brown&Mallik:2008}) as a function of photon energy $\epsilon$ with $V_{Z_{\rm eff}} = Z_{\rm eff}^2\chi$ is (all notation as in BM)

\begin{eqnarray}
\nonumber
J_{\rm R_{\rm therm}}(\epsilon) = \sqrt{\frac {2\pi}{27m_e}}\frac {64r_e^2\chi^2}{\alpha}  \frac {2n_p^2AL}{\epsilon (kT)^{3/2}} \sum_{Z_{\rm eff}} \sum_{n \geq n_{\rm min}} p_n \zeta_{\rm RZ_{\rm eff}} \frac{1}{n^3} \\ \nonumber \times ~ \exp \left(\frac{V_{\rm Z_{\rm eff}}/n^2 - \epsilon}{kT} \right) ~~;~~  \epsilon \geq V_{\rm Z_{\rm eff}}/n^2 \\
\times ~~ 0 ~~;~~ \epsilon < V_{\rm Z_{\rm eff}}/n^2.
\label{thermal}
\end{eqnarray}

\noindent
The revised thin-target non-thermal recombination expression (replacing Eq. (B.2) of \cite{Brown&Mallik:2008}) is:

\begin{eqnarray}
\nonumber J_{\rm R_{\rm thin}}(\epsilon) = (\delta -1) \frac {32 \pi}{\sqrt{3}\alpha} \frac {r_e^2 \chi^2}{\epsilon} \frac {2n_p ALF_c}{E_c^2} \sum_{Z_{\rm eff}} \sum_{n \geq n_{\rm min}} p_n \zeta_{\rm RZ_{\rm eff}} \frac{1}{n^3} \\ \nonumber \times ~ \left[\frac{\epsilon - V_{\rm Z_{\rm eff}}/n^2}{E_c} \right]^{-\delta - 1} ~~ ; ~~ \epsilon \geq E_c + V_{\rm Z_{\rm eff}}/n^2 \\
\times ~~ 0 ~~;~~ \epsilon < E_c + V_{\rm Z_{\rm eff}}/n^2
\label{thin}
\end{eqnarray}

\noindent
for the cut-off power-law $F(E)$. We also give the corrected thin target expression for general $F(E)$ in Eq. \ref{gen} to enable readers to extend our results:

\begin{equation}
J_{\rm thin}^{F(E)} (\epsilon)= \frac{32\pi
r_e^2\chi^2n_pV}{\epsilon}\sum_Z\sum_{n\geq n_{\rm min}} p_n \frac{A_Z
Z^4}{n^3}\frac{F(E-Z^2\chi/n^2)}{E-Z^2\chi/n^2}.
\label{gen}
\end{equation}

The thick-target NTR expression (revised from Eq. (B.12) of \cite{Brown&Mallik:2008}) is

\begin{eqnarray}
\nonumber J_{\rm R_{\rm thick}}(\epsilon) = \frac{32 \pi r_e^2}{3\sqrt{3} \alpha} \frac{\chi^2 {\cal F}_{oc}}{K \epsilon} \sum_{Z_{\rm eff}} \sum_{n \geq n_{\rm min}} p_n \zeta_{\rm RZ_{\rm eff}} \frac{1}{n^3} \\
\nonumber \times ~ \left[\frac {\epsilon - V_{\rm Z_{\rm eff}}/n^2}{E_{oc}} \right]^{-\delta_o + 1} ~~;~~ \epsilon \geq E_{oc} + V_{\rm Z_{\rm eff}}/n^2 \\
\nonumber \times ~ \left[\frac {E_{oc} - V_{\rm Z_{\rm eff}}/n^2}{E_{oc}} \right]^{-\delta_o + 1} ~~;~~ V_{\rm Z_{\rm eff}}/n^2 < \epsilon < E_{oc} + V_{\rm Z_{\rm eff}}/n^2 \\ \times ~~ 0~~;~~ \epsilon \leq V_{\rm Z_{\rm eff}}/n^2.
\label{thick}
\end{eqnarray}

Eqs. (\ref{thin}) and (\ref{thick}) also correct several typos in expressions (B.2) and (B.12) of BM respectively.  Thick target results are for the total emission rates over continuously injected electron collisional lifetimes.

 Although all $n$ values should strictly be included for all ions, in practice only the recombination to the lowest available level is significant, because the rates are $\propto 1/n^3$. Hence, in our calculations, we use only $n=n_{\rm min}=1$ for all elements up to Ca, while from Ca onwards, we have $n=n_{\rm min}=2$ for the appropriate ions, e.g. Fe 23+, Fe 24+ or Ni 25+. With very high temperatures Fe 25+ and 26+ also become important. For these species, $n_{\rm min}=1$.

The ionisation equilbrium used was a fit to the standard steady state
coronal collisional ionisation, optical depths being negligible.
In Table \ref{tab1}, we have listed the abundances and $\zeta_{RZ}$ for
Fe 25+ and Fe 26+ \citep{Arnaud&Raymond:1992} for a set of four
temperatures, and these can be seen in addition to the relevant
values in Tables 1 and 2 of \cite{Brown&Mallik:2008}, where these values are listed for fully ionised Fe (Fe 26+) at $T\gg 100$ MK and for the most abundant Fe species at 20 MK. Here we have also
included the ratio of non-thermal free-bound (due to only Fe 25+ and 26+ respectively) to total free-free
flux, $R = J_{NTR_{\rm Fe26+,25+}}/J_{NTB}$ (from Eq. 2 of \cite{Brown&Mallik:2009}), at $\epsilon=20$ keV and $\delta=5$ to
illustrate at what temperatures Fe 25+ and Fe 26+ start becoming
significant contributors to the NTR flux.

Note that we use the variable $f_c$ to define the fraction of total electrons that are accelerated, i.e. $f_c = n_c/n_p$, where $n_c$ is the density of non-thermal electrons.

\begin{table*}
\centering
\caption{Significance of Fe 25+ and 26+ at 4 different temperatures for
$\epsilon=20$ keV and $\delta=5$.}
\vskip 0.2cm
\begin{tabular}{ccccccc}
\hline\smallskip
$T$ & $A_Z$ & $A_Z$ & $\zeta_{RZ}$ & $\zeta_{RZ}$ & $R$ & $R$
\\
(MK) & (Fe 25+) & (Fe 26+) & (Fe 25+) & (Fe 26+) & (Fe 25+) & (Fe 26+)
\\
\hline
20 & $2.4 \times 10^{-7}$ & $2.5 \times 10^{-10}$ & $4.7 \times
10^{-2}$ & $1.1 \times 10^{-4}$ & $1.0 \times 10^{-2}$ & $3.4 \times
10^{-5}$\\
30 & $3.5 \times 10^{-6}$ & $4.5 \times 10^{-8}$ & 0.68 & $2.1 \times
10^{-2}$ & 0.15 & $6.5 \times 10^{-3}$\\
40 & $6.7 \times 10^{-6}$ & $2.8 \times 10^{-7}$ & 1.3 & 0.13 & 0.28 &
$4.0 \times 10^{-2}$ \\
50 & $1.7 \times 10^{-5}$ & $1.3 \times 10^{-6}$ & 3.3 & 0.59 & 0.70 &
0.18 \\
\hline
\end{tabular}
\label{tab1}
\end{table*}

\section{Results}

\begin{figure}
\centering
\includegraphics[width=0.5\textwidth]{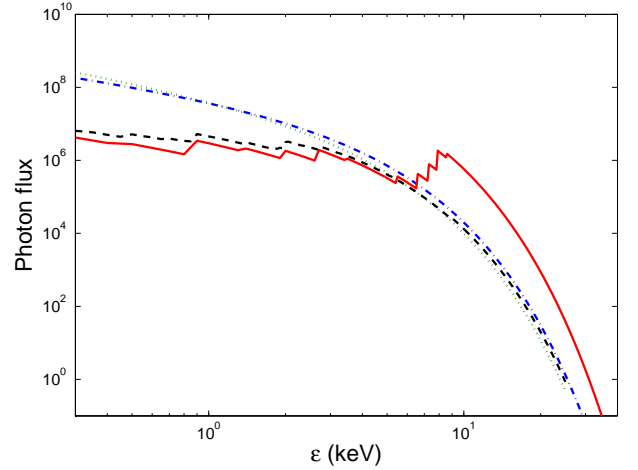}
\caption{The original BM thermal model spectra (dot-dash blue is TB; solid red is TR) compared with Chianti's (dotted green for TB; dashed black for TR) for $T=20$ MK. Courtesy E. Landi for the Chianti spectra.}
\label{chianticomp}
\end{figure}

\begin{figure}
\centering
\includegraphics[width=0.5\textwidth]{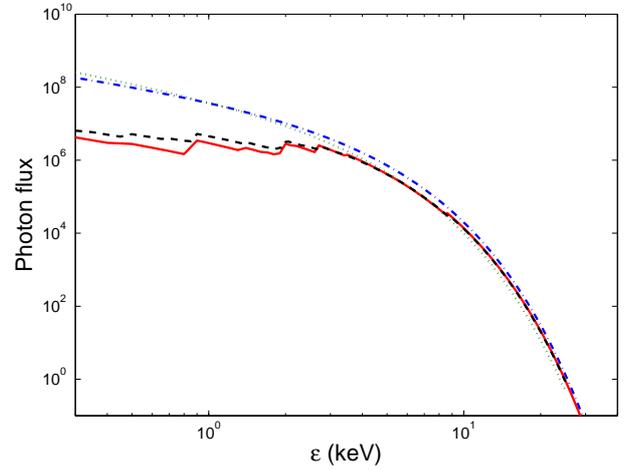}
\caption{The revised thermal model spectra compared with the same Chianti spectra for 20 MK; the colour/linestyle-coding is the same as in Fig. \ref{chianticomp}}
\label{comp1}
\end{figure}

\begin{figure}
\centering
\includegraphics[width=0.5\textwidth]{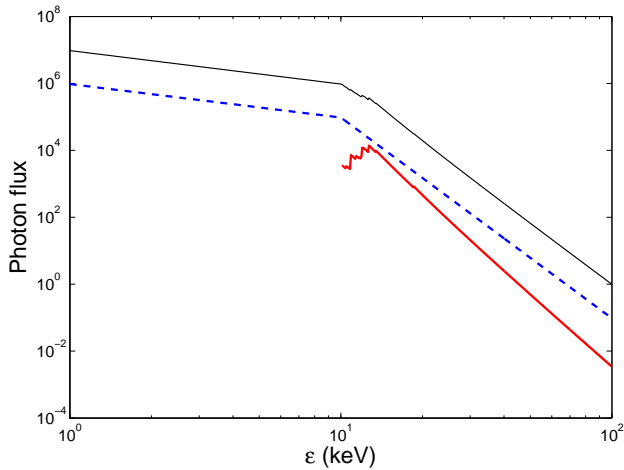}
\caption{NTB (dashed blue) and NTR (solid red) spectra from the revised thin-target model with the solid black curve denoting the total non-thermal flux multiplied by 10. $E_c = 10$ keV, $\delta = 5$ and $T=20$ MK.}
\label{comp2}
\end{figure}

\begin{figure}
\centering
\includegraphics[width=0.5\textwidth]{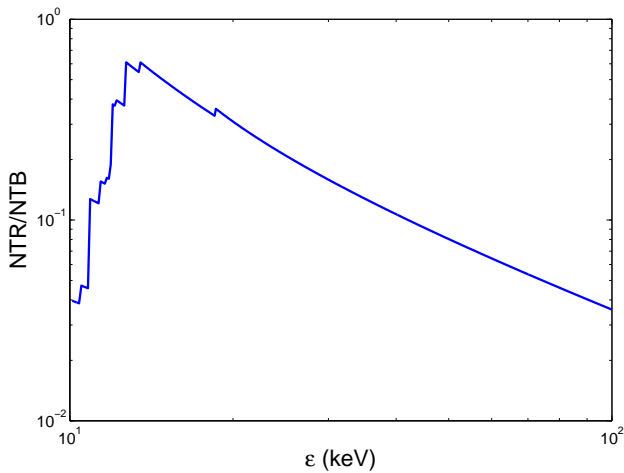}
\caption{Thin-target NTR:NTB ratio for the same flare parameters as in Fig. \ref{comp2}}
\label{comp3}
\end{figure}

\begin{figure}
\centering
\includegraphics[width=0.5\textwidth]{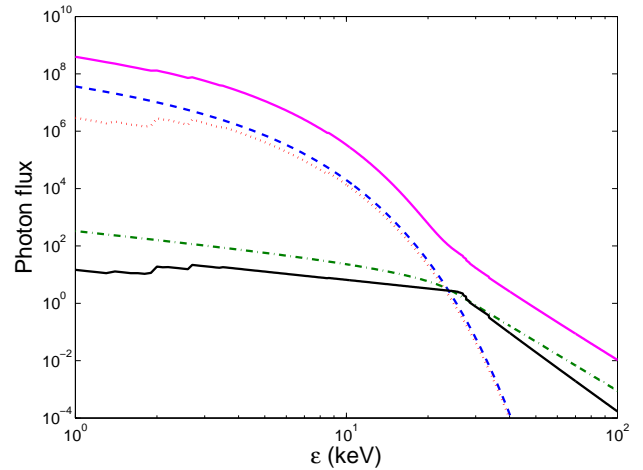}
\caption{Our revised model thick-target plot spectra for the 2002 April 14 event. The dashed blue curve is TB, dotted red is TR, dot-dash green is NTB and solid black is NTR. The solid magenta curve is the total flux multiplied by 10; small TR and NTR edges are clearly visible but much less distinct from what was predicted in Fig. 5 of \cite{Brown&Mallik:2008}}
\label{comp4}
\end{figure}

\begin{figure}
\centering
\subfigure{\includegraphics[width=0.4\textwidth]{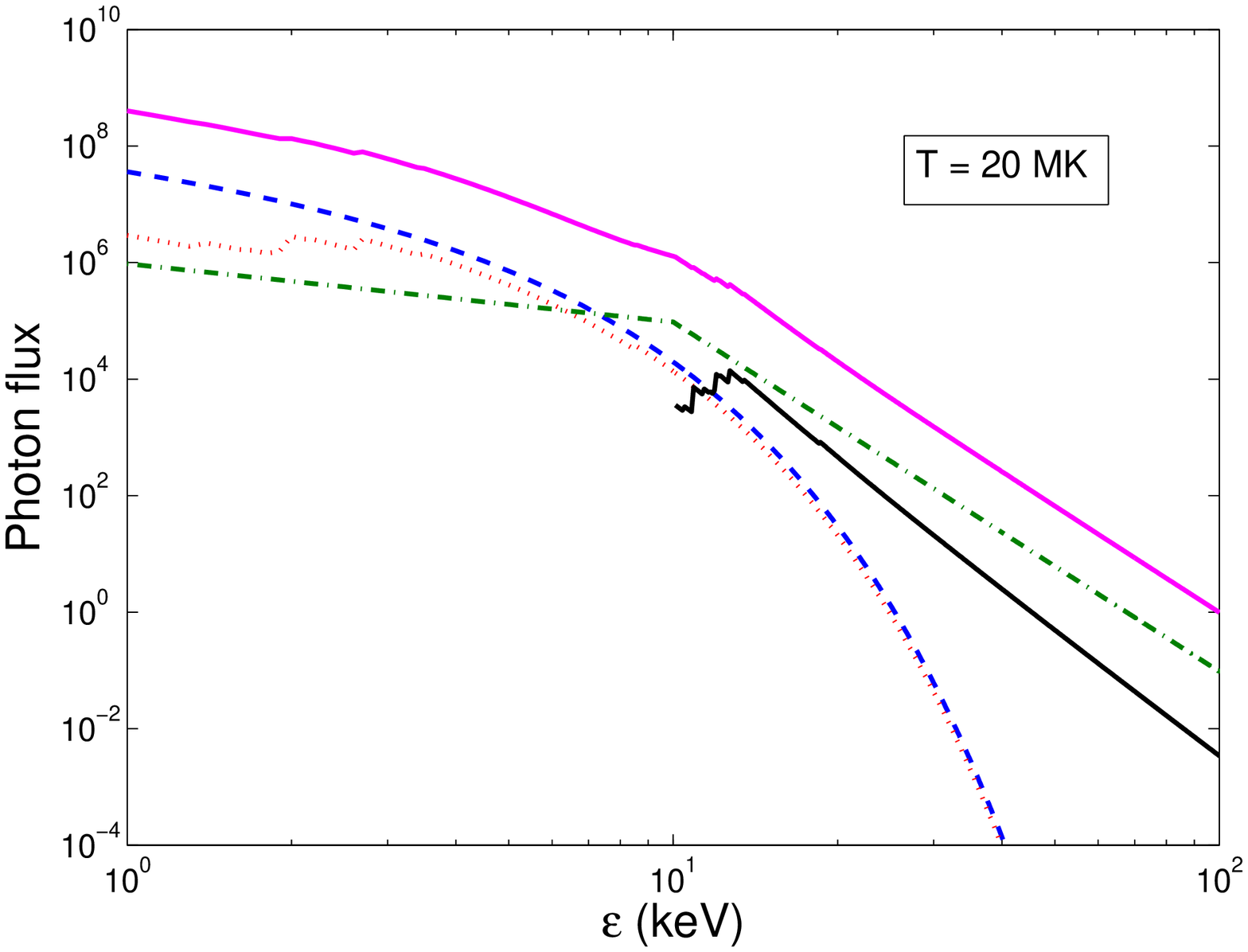}}
\subfigure{\includegraphics[width=0.4\textwidth]{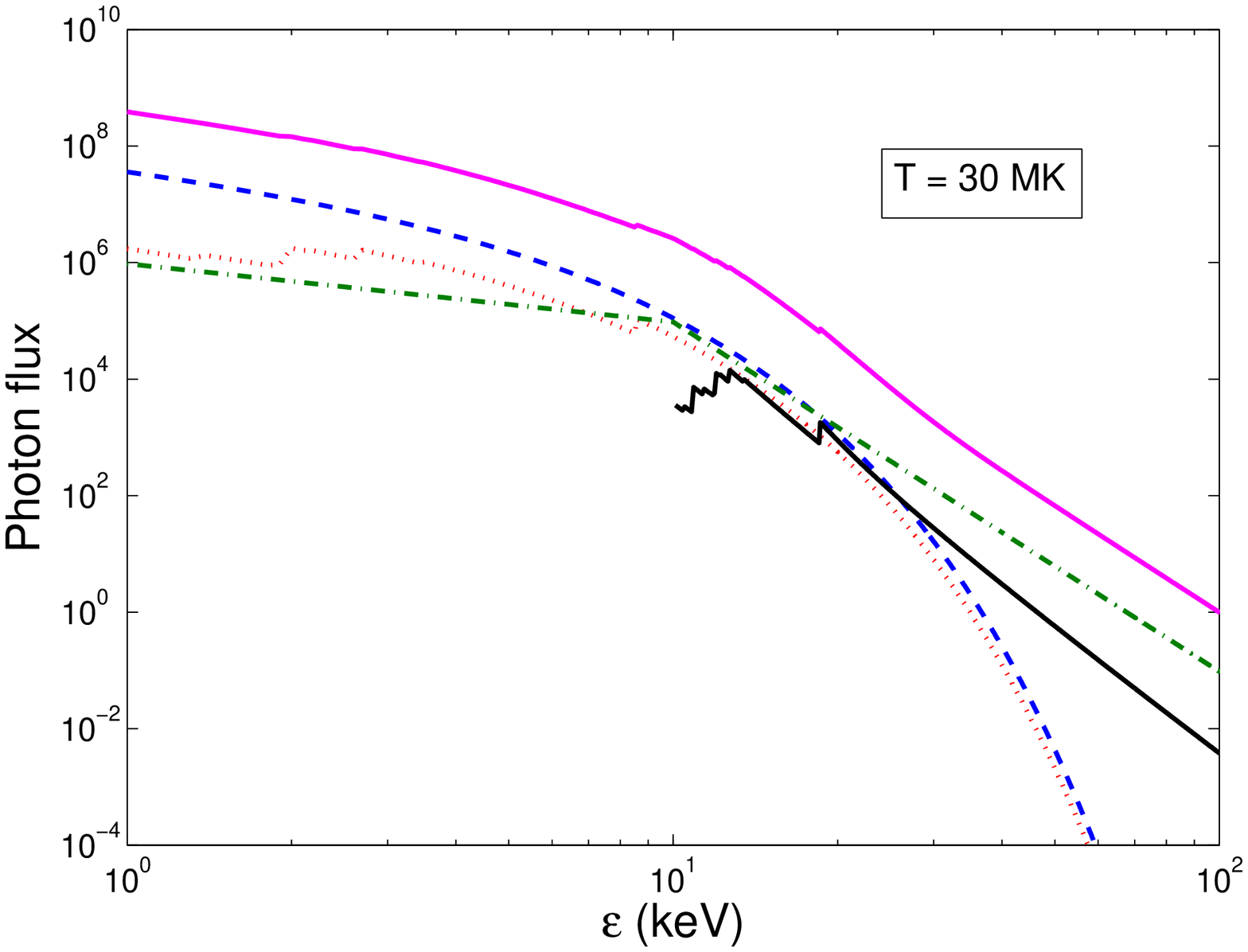}}
\subfigure{\includegraphics[width=0.4\textwidth]{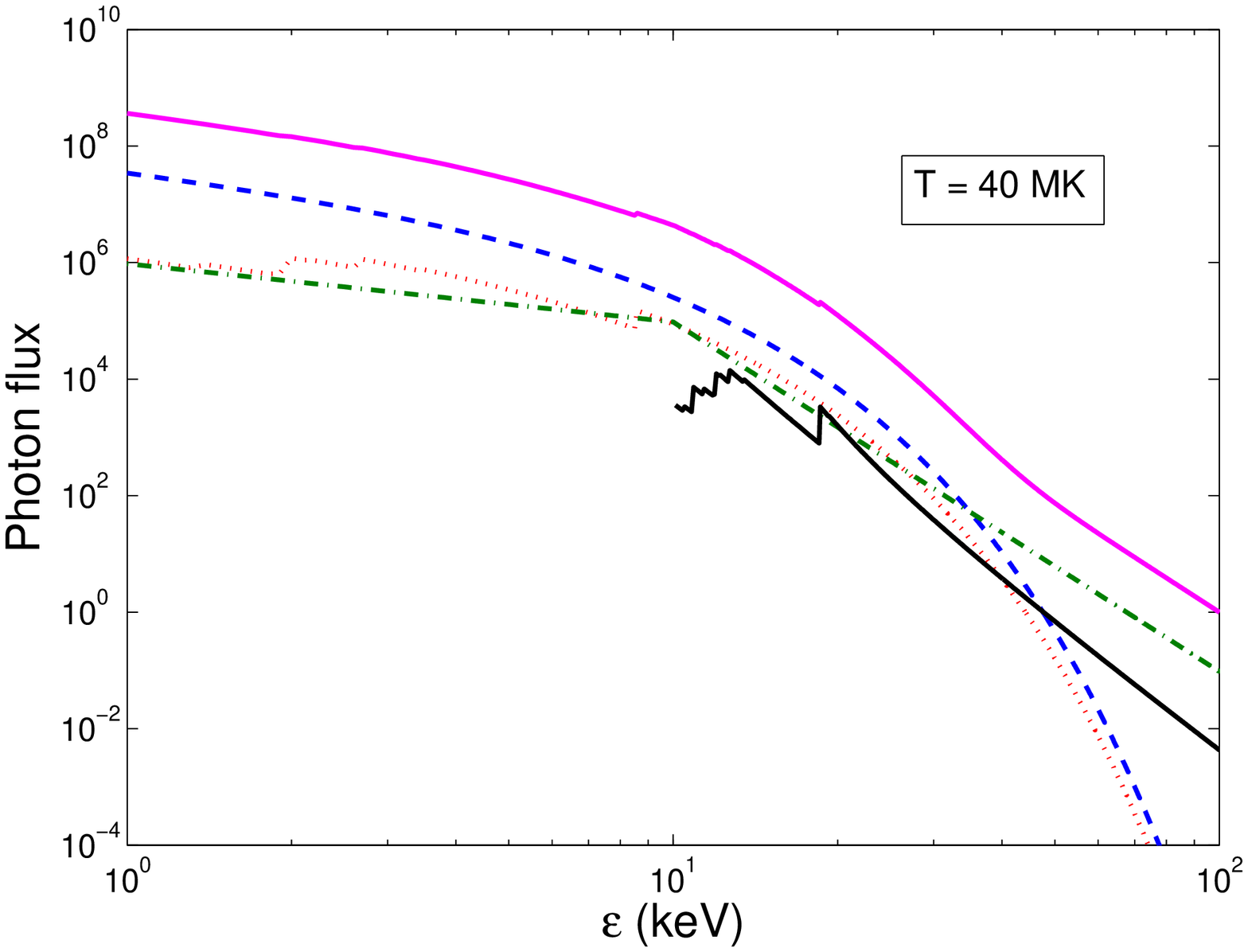}}
\subfigure{\includegraphics[width=0.4\textwidth]{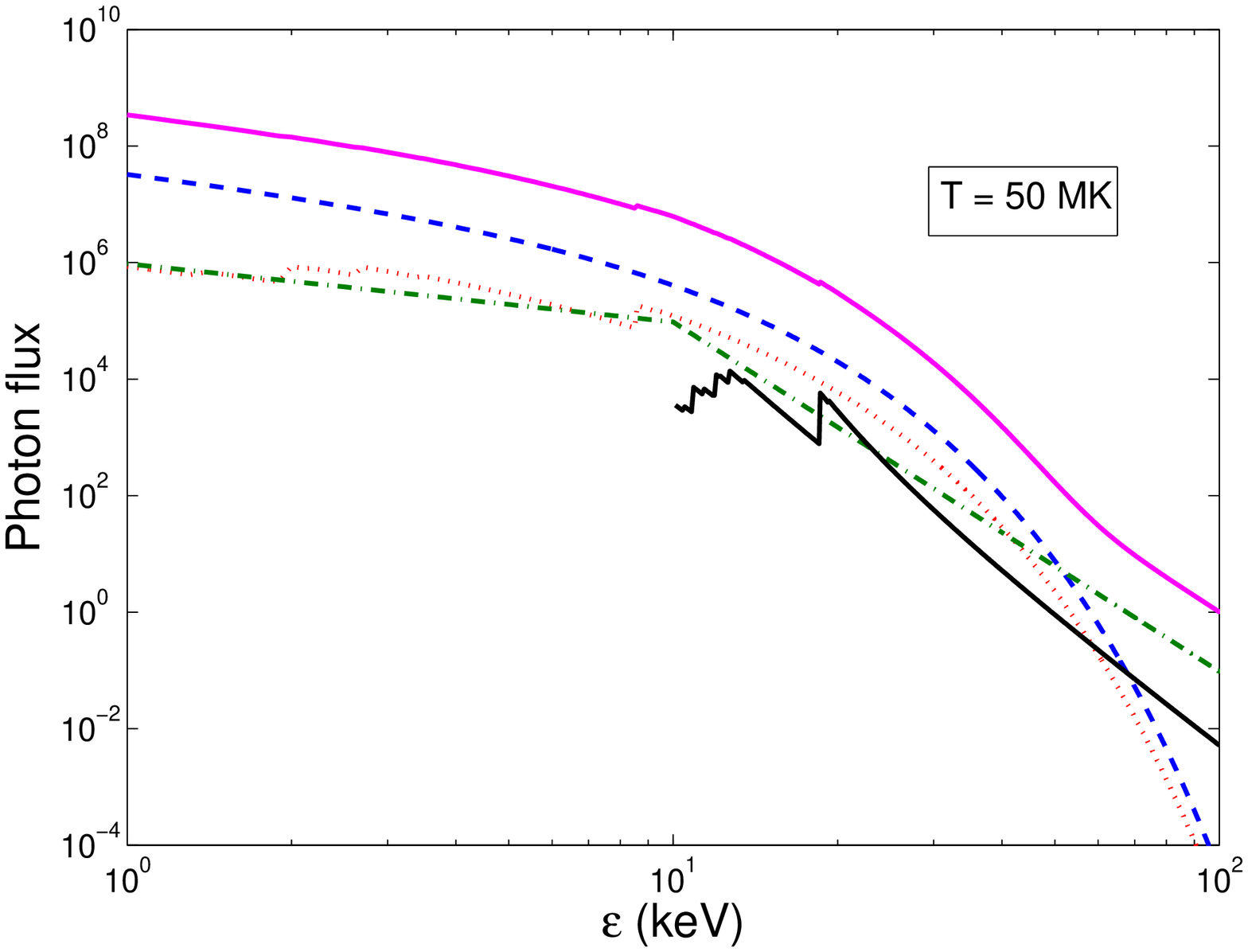}}
\caption{The revised thin-target spectra for 4 different temperatures with $f_c = 0.1$. Note that the non-thermal emission
measure, $EM_c = f_cEM$, where $EM=2ALn_p^2$ is the thermal emission
measure. The colour/linestyle-coding is the same as in Fig. \ref{comp4}}
\label{comp6}
\end{figure}

\begin{figure}
\centering
\includegraphics[width=0.5\textwidth]{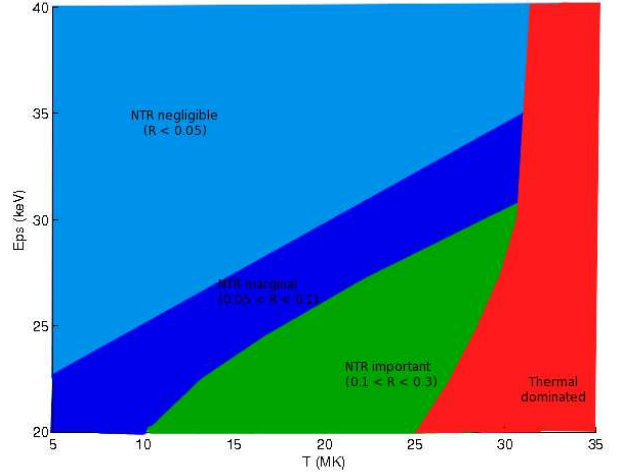}
\caption{Regime plot with our revised model showing the relevant areas of importance in the $(\epsilon, T)$ domain for $E_c = 10$ keV, $\delta = 5$ and $f_c = 0.01$}
\label{comp7}
\end{figure}

\begin{figure}
\centering
\includegraphics[width=0.5\textwidth]{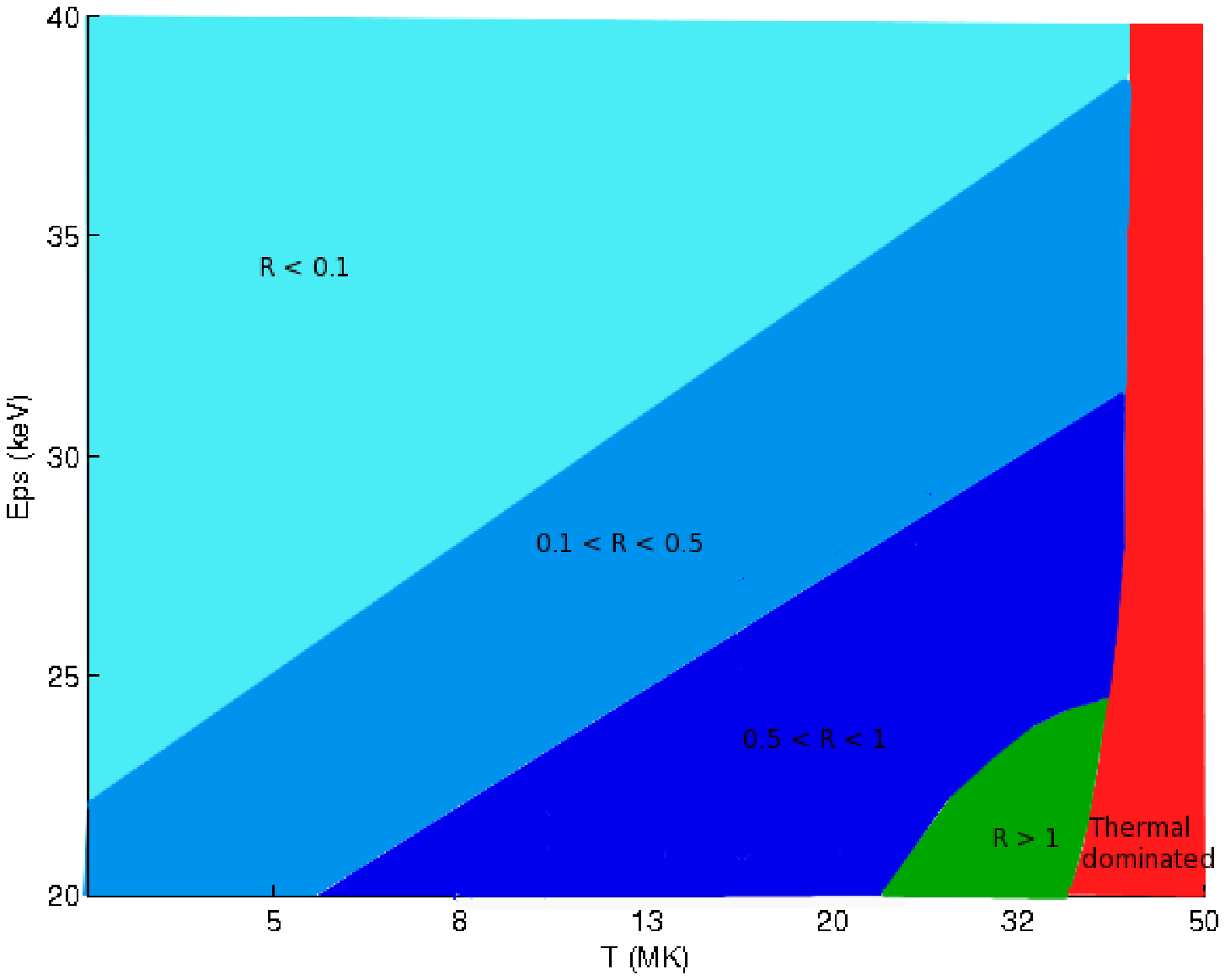}
\caption{Regime plot with our revised model showing the relevant areas of importance in the $(\epsilon, T)$ domain for $E_c = 10$ keV, $\delta = 5$ and $f_c = 1$. This shows that NTR may still be a dominant source of HXRs when $f_c$ approaches unity.}
\label{comp11}
\end{figure}

Some of the plots from BM have been reproduced here to show the essential changes in our results. We have also added some new plots to show interesting new results. We have not reproduced all the plots as that would be repetitive but, from these sample plots, it should be clear what the changes are and what they imply. Throughout this section we have used a resolution of 0.1 keV bins, unlike in BM, where primarily 1 keV bins were used. We do this so that the finer details can be appreciated.

Firstly, Fig. \ref{comp1} shows our thermal spectra compared with those of Chianti at 20 MK. It is clear here that the major BM discrepancies  from the plot in Fig. \ref{chianticomp} have been removed. The slight differences between our model spectra and those of Chianti can be attributed to finer details such as the use of Kramers' cross-section versus more accurate calculated cross-sections. But while the thermal edges are much smaller, they can still form a significant proportion of the flux and may be detectable.

In Fig. \ref{comp2} we redo the bottom-left plot from Fig. 1 of \cite{Brown&Mallik:2008}, where $E_c = 10$ keV, $\delta = 5$ and $T=20$ MK (hot), showing only the thin-target non-thermal spectra. The main point to note here is that at no point is NTR greater than NTB, but it can still be a significant contributor to the total non-thermal flux in and around the cut-off energy, in this case 10 keV. However, the thermal emission at these energies is quite likely to limit the detectability of the non-thermal edges for these temperatures. Nevertheless, it is important to note that all the major edges occur at around 2-4 keV (or 2-4 keV above the cut-off in the case of NTR thin-target) because our revised $V_Z/n^2$ value for the important species of Fe, namely Fe 23+ and Fe 24+ in this case, is about 2-3 keV. So these important Fe edges coexist with edges from other elements like Si and O that have significant values of $\zeta_{RZ}$ and whose $V_Z =$ 3-4 keV. These are almost fully ionised, so $n_{\rm min}=1$ for them.

Fig. \ref{comp3} shows the non-thermal thin target NTR:NTB ratio for the same parameters and can be contrasted with the bottom-left plot of Fig. 2 of \cite{Brown&Mallik:2008}. These plots are virtually identical in shapes, though the revised plot has smaller values for the ratio, as expected.

Fig. \ref{comp4} is the corresponding modified thick target plot of Figure 5 from \cite{Brown&Mallik:2008}, and is our predicted spectrum for the 2002 April 14 flare parameters. Here too the non-thermal edge is comparable to the total non-thermal flux at the cut-off energy of 20 keV and might be detectable in observed spectra. The fact that NTR can be comparable to NTB for these particular real flare parameters is of particular significance. Also, the sharp nature of the observed edges can have a large effect on spectral inversions, an issue we are currently investigating.

Since Fe 25+ and Fe 26+ start becoming important only for temperatures above 30 MK (Table \ref{tab1}), we decided to look at our model spectra for each of the four different temperatures while keeping $E_c = 10$ keV and $\delta = 5$ fixed. With $f_c = 0.1$, we plot results in Fig. \ref{comp6} for the thin-target case. Although the Fe 25+/26+ edge becomes more prominent for the higher temperatures and NTR can be greater than NTB at these edges, higher temperatures also mean an increased thermal emission (where TR becomes less and less important than TB), which tends to hide these non-thermal features. Hence, if NTR is to be observed, the optimum temperature would be about 20-30 MK. The same temperature range is also optimum to observe TR edges, as is validated by \cite{Landi:2007}, who also predicts that TR starts becoming less significant for $T>25$ MK.

The surface plot in Fig. \ref{comp7} is analogous to Fig. 2 of \cite{Brown&Mallik:2009}, showing the regions of relevant importance. It is evident that NTR is never dominant but still can be several 10s of percent of the NTB flux. Once again, increasing $f_c$ to 0.1 pushes the thermal-dominated line by about 5 MK to the right, but of course never changes the NTR:NTB ratio. However, in the $(\epsilon, T)$ range of 20-30 keV and 20-30 MK respectively, NTR can still be significant enough for its inclusion in spectral analysis packages to be essential, moreso when $f_c$ is even larger as that increases the region of importance. As an extreme case, we produce Fig. \ref{comp11}, where $f_c\rightarrow 1$ (as in \cite{Krucker:2008c}). Thus, when $f_c$ is large, NTR dominated regions still exist in $(T,\epsilon)$, albeit at rather high $T$ and over a modest $\epsilon$ range. Nonetheless, $R$ is substantial ($>0.5$) for a large $(\epsilon, T)$ regime of 15-40 MK and energy range 20-35 keV, which is crucial when considering albedo spectral corrections \citep{Alexander&Brown:2002} and also electron number and energy budgets. Even if $f_c$, $T$ and $\delta$ are not {\it so} high, NTR can still be up to a 20-40\% effect.

(Note that all the other parameters for all these plots have been maintained throughout, viz.: density $n_p = 1.5 \times 10^{11}$ cm$^{-3}$, flare loop cross-section area $A = 2 \times 10^{17}$ cm$^2$ and loop half-length $L = 2.5 \times 10^9$ cm, mirroring those observed for the 2002 April 14 event \citep{Veronig&Brown:2004}).

\section{Conclusions and implications}

The upshot of our revised model is that although NTR is (contrary to BM) seldom dominant over NTB {\it and} the total thermal emission, it can still be substantial enough for its inclusion to be essential. The recombination edges clearly seen in our model spectra (Fig. \ref{comp4}), though small, could have a large impact on electron spectra inferred by spectral inversion since this involves differentiation of the photon spectrum. In extreme parameter regimes, NTR can be a substantial or dominant source of HXRs and so a large consideration in electron number and energy budgets and hence acceleration efficiency problems. However, for the majority of hot HXR sources, NTR is unlikely to be dominant but still significant enough to demand consideration.

Another enhancement to the NTR contribution is the possible enrichment of Fe abundance in the corona compared to photospheric levels by a factor even higher than the factor of 3 we have used based on \cite{Feldman:1992}. Factors of $\sim$ 10 or more have been suggested by \cite{Feldman:2004}, which would increase our NTR predictions by another factor of 3. In that case even for `normal' flare parameters, NTR would start pushing the NTR-dominant boundary. Clearly, much work needs to be done to see if the First Ionisation Potential (FIP) effect really does enrich low FIP elements, such as Fe, to the levels mentioned by \cite{Feldman:2004}.

Further improvements to our model can be made by using more accurate recombination cross-sections instead of Kramers. But our research has revealed that Kramers' cross-section is not a bad approximation and is accurate to within 20\% of calculated cross-sections for highly-charged ions. Another refinement in our model would be to include recombination onto 2 or 3 higher levels apart from the lowest available level of the ion. This too would make our model more accurate by a few percent at the most. In any case, it will be necessary to include NTR in spectral analysis packages such as SolarSoft to get reliable inferences from flare data.

\acknowledgements
We gratefully acknowledge the financial support of a UK STFC Rolling Grant
(JCB), of a Dorothy Hodgkin's Scholarship (PCVM), an ISSI Grant (JCB), and an STFC Grant (NRB). Discussions with E.P. Kontar, R.A. Schwartz, and E. Landi were invaluable in drawing attention to the fact that there was likely something amiss with the BM results, which we have addressed in this paper.

\bibliographystyle{aa}
\bibliography{pcvm}

\end{document}